\documentclass[aps,prl,preprint,groupedaddress,superscriptaddress]{revtex4-1}
\usepackage[usenames]{color}
\usepackage{natbib}
\usepackage{graphicx}
\usepackage{float}
\begin{document}


\title{Vacancy in graphene: insight on magnetic properties from theoretical modeling}

\author{A. M. Valencia}
\email[]{valencia@if.usp.br}
\affiliation{Instituto de F\'{\i}sica, Universidade de S\~ao Paulo, CEP 66318, 05315-970, S\~ao Paulo-SP, Brazil}
\author{M. J. Caldas}
\affiliation{Instituto de F\'{\i}sica, Universidade de S\~ao Paulo, CEP 66318, 05315-970, S\~ao Paulo-SP, Brazil}

\date{\today}

\begin{abstract}
	Magnetic properties of a single vacancy in graphene is a relevant and still unsolved problem. The experimental results point to a clearly detectable magnetic defect state at the Fermi energy, while several calculations based on density functional theory (DFT) yield widely varying results for the magnetic moment, in the range of $\mu=1.04-2.0$ $\mu_{B}$. We present a multi-tool \textit{ab initio} theoretical study of the same defect, using two simulation protocols for a defect in a crystal (cluster and periodic boundary conditions) and different DFT functionals - bare and hybrid DFT, mixing a fraction of exact Hartree-Fock exchange (XC). Our main conclusions are two-fold: First, we find that due to the $\pi$-character of the Fermi-energy states of graphene, inclusion of XC is crucial and for a single isolated vacancy we can predict an integer magnetic moment $\mu=2\mu_{B}$. Second, we find that due to the specific symmetry of the graphene lattice, periodic arrays of single vacancies may provide interesting diffuse spin-spin interactions.
\end{abstract}

\maketitle

A single vacancy is the simplest intrinsic defect in a crystal, and for covalently-bonded crystals a strong effect on electronic and magnetic properties is expected. The dangling-bonds left on the neighbor atoms usually lead to local symmetry-breaking, through a Jahn-Teller rearrangement, and e.g. in 3D semiconductors we find a localized state and deep gap levels. Graphene on the other hand has notable 2D properties with the covalent bonding introducing two intrinsically different state types,  $\sigma$ and $\pi$, these last relevant for the Fermi-energy and Dirac point properties. The $\pi$-states are diffuse in the 2D planar ($x,y$) directions, but very localized on the $z$-direction with an in-plane node. As such, long range 2D electron-electron interaction is enhanced. In addition, the hexagonal structure with two sublattices creates for the $\pi$ states the special band structure with the Dirac point. We might thus expect special properties also for the vacancy in graphene. Experimental studies find a clear symmetry for the defect, and in particular from scanning tunneling microscopy~\cite{uged+10prl,zhan+16prl} it is found also that the defect level is resonant at the Dirac point, and induces magnetism~\cite{zhan+16prl}.

A number of theoretical studies of the electronic and magnetic properties of the vacancy in graphene have been reported in the past decade \cite{pala+08prb,pere+06prl,caza+12arxiv,mira+16prb,elba+03prb,leht+04prl,ma+04njp,yazy&helm07prb,pala&yndu12prb,casa+13prb, lin15jpcc,mene&capa15jpcm,rodr+16carbon,padm&nand16prb,zhan+16prl}. In particular, first-principles calculations based on density functional theory (DFT) ~\cite{elba+03prb,leht+04prl,ma+04njp,yazy&helm07prb,pala&yndu12prb,casa+13prb, lin15jpcc,mene&capa15jpcm,rodr+16carbon,padm&nand16prb,zhan+16prl}  yielded widely varying results for the magnetic moment in the range of $1.04-2.0$ $\mu_{B}$. For instance, Palacios and Yndur\'ain~\cite{pala&yndu12prb} found that the magnetization decreases with decreasing defect density, tending to $1.0 \mu_{B}$ in the low-density limit, in contrast to results reported by Yazyev and Helm~\cite{yazy&helm07prb} where the magnetization increases from $1.15 \mu_{B}$ to $\sim 1.5 \mu_{B}$ with decreasing density, results that highlight the possible dependence of magnetic moment with defect-defect interaction.

Regarding this last point, two typical approaches can be used for the simulation: model clusters, which are assumed to resemble the defect environment in the bulk, or periodic boundary conditions based on the choice of \textit{supercells} (SC). In the cluster model we must be careful about defect interaction with cluster edge states, which in the case of graphene can be critical~\cite{fern&pala07prl,wang+08nl}. As for the SC modeling, we must remember that we will study defects periodically arranged \cite{garc10prb}, that is, we study an \textit{array of defects} that may induce spurious defect interactions.

\begin{figure*}
	\centering
	\includegraphics[width=0.9\textwidth]{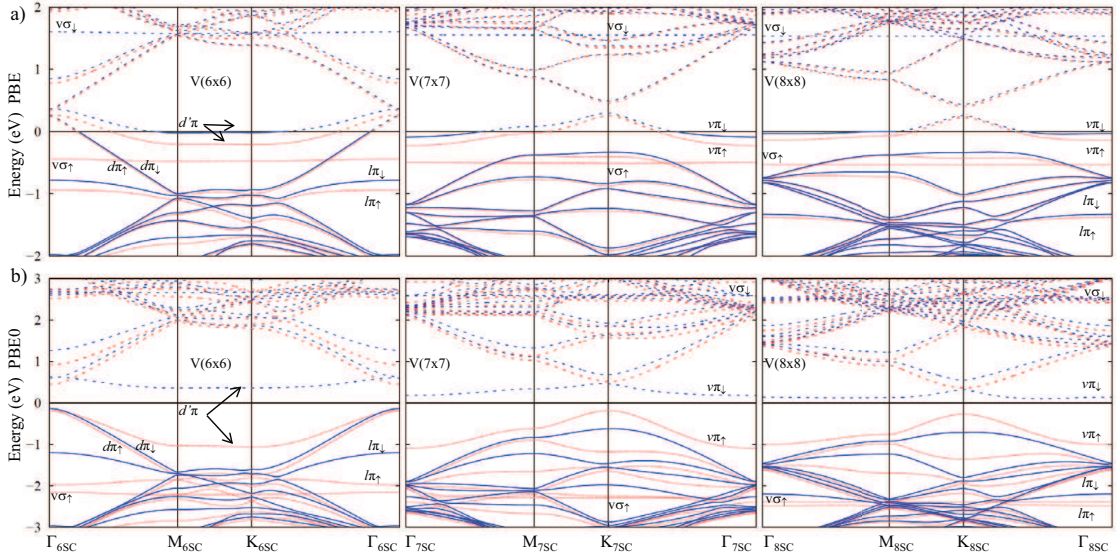}
	\caption{(Color online)Band structure for the vacancy defect in the region near the Fermi energy, results from spin-polarized PBE (a) and PBE0 (b). Solid (dotted) lines indicate occupied (unoccupied) states. Periodic conditions with symmetric SCs ($6\times6$), ($7\times7$) and ($8\times8$). Energies aligned to the Fermi energy of the perfect crystal by the C-$1s^{2}$ average energy.}
	\label{v_level_band_PBE_PBE0_sp}
\end{figure*}

Concerning the defect-edge interactions and focusing on the $\pi$-states, when we have zig-zag edges we can (depending on the cluster symmetry) bring in Lieb's imbalance states~\cite{lieb89prl} that will group at the Fermi energy. These states are not realistic concerning the modeling of infinite graphene (no Lieb's imbalance). In the case of SCs we have for graphene three different symmetrical ($N\times N$) families, as shown in Refs. \onlinecite{aike+15prb,ding+11prb}, namely ($3n\times3n$), ($3n-1\times3n-1$) and ($3n+1\times3n+1$), where $n$ is an integer number. For the  $3n$ family, there occurs a folding of the $K-K'$ points onto the $\Gamma$-point of the SC Brillouin zone, that is, we will have degenerate, fully delocalized $\pi$-character states of different original symmetry crossing the Fermi energy at the SC $\Gamma$-point. These delocalized states interfere with localized defect states, through the long-range interaction property of the $\Gamma$-point, and is avoided when we adopt either one of the other families. Still for supercells, due to the $\pi$-symmetry of the relevant states at the Fermi region, we also have to take into account the possibility of long-range interaction between defects coming from parity in the zig-zag direction, as will be seen here.

In this work we adopt both the cluster approach, choosing hexagonal clusters with arm-chair and zig-zag edges, and periodic conditions with symmetrical cells from the different families:($3n\times3n$)($6\times6$), ($3n+1\times3n+1$)($7\times7$) and ($3n-1\times3n-1$)($8\times8$). We use semi-local DFT~\cite{perd+96prl} and hybrid DFT including a fraction $\alpha$ of XC~\cite{perd+96jcp}, in which $\alpha$ is chosen to reproduce the properties of perfect graphene in the Fermi energy region~\cite{pinh+15prb}. We find that, for the isolated vacancy defect, we can predict it introduces a magnetic moment of $2\mu_B$. Moreover, we find that periodic arrays of the defect can bring in interesting long-range spin dispersion effects~\cite{just+14prb,ulyb-kats15prl}.

All calculations are performed through the all-electron FHI-aims code~\cite{blum+09cpc} with spin-polarization: the code employs numeric atom-centered orbitals obtained from \textit{ab-initio} all-electron calculations for isolated atoms, and can be used at the mean-field level with finite or infinite periodic models. The use of an all-electron code allows us to align the level structure of different simulation models by the deep 1$s^{2}$ Carbon orbitals. We employ \textit{tight} integration grids and \textit{tier2} basis sets~\cite{havu+09jcp}, and the atomic positions are relaxed until the Hellmann-Feynman forces are smaller than 10$^{-3}$ eV/\AA{}, without any symmetry restriction, through the GGA functional of Perdew \textit{et al.}~\cite{perd+96prl} (PBE). For periodic cells, we use the Monkhorst-Pack~\cite{monk+76prb} ($\Gamma$-point included) scheme for Brillouin-zone sampling, with a [$6\times6\times1$] grid. The gaussian smearing is 0.01 eV. For all the systems shown here we find that the final structure is planar, and the vacancy formation energy is in the range of $7.63-7.70$ eV.

Standard DFT with local or semilocal exchange-correlation functionals is known to suffer from self-interaction errors~\cite{cohe+08science}(SIE) leading to excessive delocalization of electrons~\cite{kumm&kron08rmp}. Hybrid density functionals reduce the SIE by mixing in a fraction $\alpha$ of Hartree-Fock exchange (XC) and can in many cases significantly improve the study of electronic properties. A specific much used hybrid functional is PBEh~\cite{perd+96jcp}, a one-parameter hybrid functional based on PBE. The choice of $\alpha$ can be directed to the system studied by choosing a specific electronic property: here we are interested in the properties of states close to the Fermi energy, and we adopt a strategy similar to that suggested in Ref. \onlinecite{pinh+15prb}, by obtaining a proper value for the work function $E_W$ of graphene (mostly from experimental data in this case), and also for the Fermi velocity $v_F$ where many-body theoretical studies are already present~\cite{trev+08prl,yang+09prl}. We find that the original PBE0 functional~\cite{perd+96jcp} ($\alpha=0.25$) gives a good performance, with $E_W=4.35 eV$ and  $v_F = 1.3\times10^6 m/s$, compared to PBE ($E_W=4.24 eV$ and  $v_F = 0.98\times10^6 m/s$).

\begin{figure}
	\centering
	\includegraphics[width=0.5\textwidth]{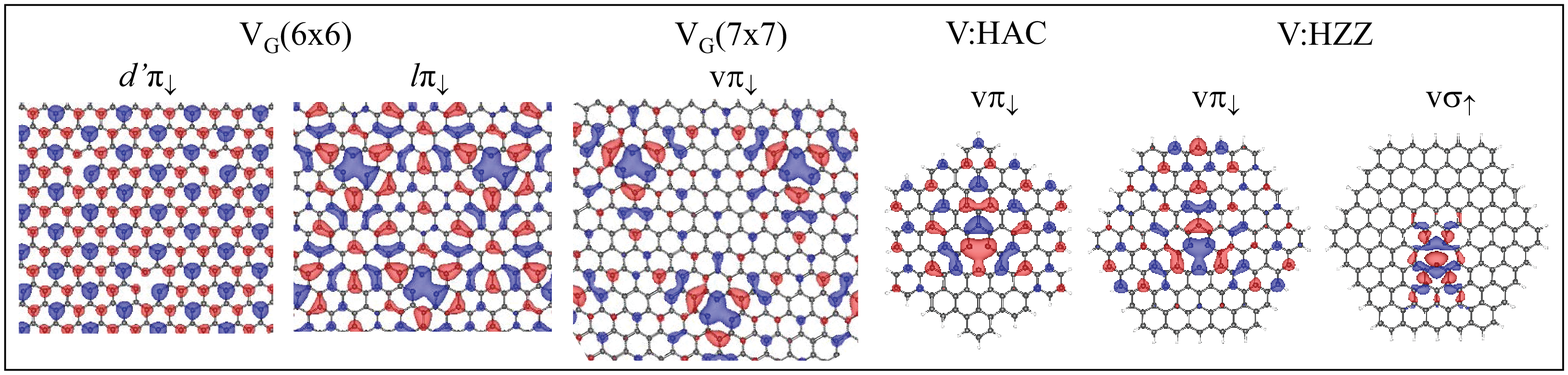}
	\caption{(Color online)Isosurfaces for the \textit{V}$\pi$, \textit{d'}$\pi$ and \textit{l}$\pi$ defect states (indicated  in Figs. \ref{v_level_band_PBE_PBE0_sp} and \ref{cluster_results}) obtained with the spin-polarized PBE functional; the \textit{V}$\sigma$ state shown here for a cluster presents very similar character in all simulations, and the \textit{l}$\pi$ state is similar in the ($8\times8$) SC. Results from PBE0 gain increased localization, maintaining the overall characteristics.}
	\label{isosurfs}
\end{figure}

We show in Fig.~\ref{v_level_band_PBE_PBE0_sp}(a) the results obtained with the PBE functional for the SC modeling.  We see first that, in all cases, we find non-integer magnetic moments $\mu_{V}=1.49\mu_{B}$ ($6\times6$); $1.30\mu_{B}$ ($7\times7$) and $1.38$ ($8\times8$), coming from the crossing of bands at the Fermi energy. We stress however that the picture is qualitatively different when moving from the ($6\times6$) to the other supercells. In the first case the bands crossing the Fermi energy, seen also in previous works \cite{pala&yndu12prb,mene&capa15jpcm,rodr+16carbon,padm&nand16prb}, are rather delocalized,  not strictly defect-localized states. Indeed, while for the disruption of the $\sigma$-states we see quite localized defect states (flat bands) \textit{{V}$\sigma$} at around $-0.5$eV, with a sizeable spin-splitting ($\sim 2.1$ eV), the effect on the $\pi$-electrons for the ($6\times6$) SC is more spread-out and affects a numbers of states (or bands), in particular the folded bands from the (K, K') unit-cell points that we call here $d_\pi$ and $d'_\pi$. The $\pi$-states characteristic of the vacancy $l_\pi$ are now quasi-localized, and are in this case affected by the parity of the SC, interacting through the zig-zag connection as shown in Fig. \ref{isosurfs}; their influence on the final spin is not direct (both up- and down-spin states fully occupied) however the impact on the spin-density is seen. Looking now at our results for the ($7\times7$) SC, free from the symmetry-folding problems and parity-connection, we see that the defect-related band $V_\pi$ is the one causing the final (non-integer) magnetic moment. Again for the ($8\times8$) SC there is no symmetry-folding, thus the $V_\pi$ state is the one crossing the Fermi energy, however we have parity connection and the $l_\pi$ states also contribute to the final magnetic moment. In summary, the defect-related states in these SCs show not only different total magnetic moment, but also very different character.

\begin{figure}
	\centering
	\includegraphics[width=0.5\textwidth]{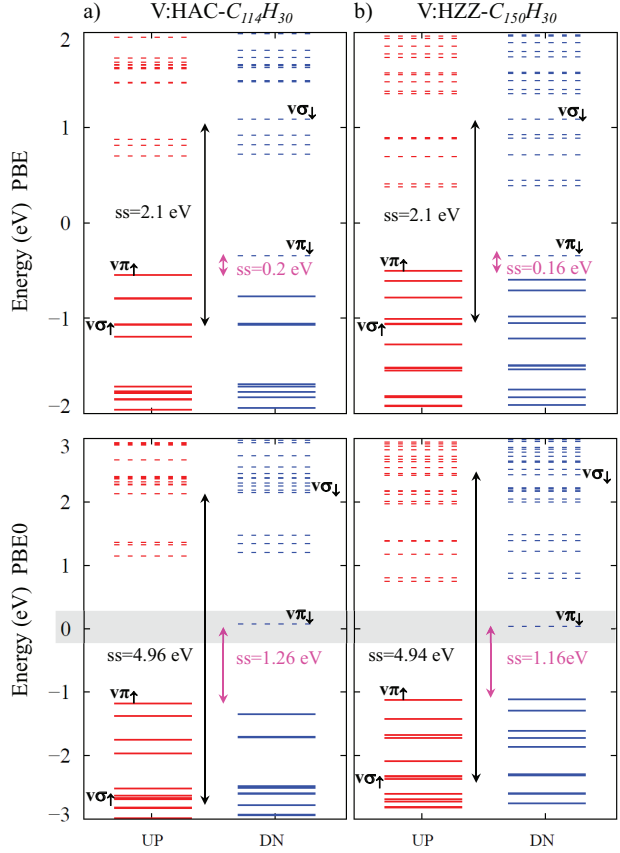}
	\caption{(Color online)Electronic energy levels for the vacancy defect in the cluster models HAC-$C_{114}H_{30}$ (left) and HZZ-$C_{150}H_{30}$ (right), in the region near the Fermi energy. Results from spin-polarized PBE (top) and PBE0 (down) functionals. Solid (dotted) lines indicate occupied (unoccupied) states. Energies aligned to the Fermi energy of the perfect crystal by the C-$1s^{2}$ average energy.}
	\label{cluster_results}
\end{figure}

We now turn to the results from the two cluster models, shown in Fig. \ref{cluster_results}; the $\sigma-\pi$ character of graphene allows us to use Hydrogen-saturation, and we adopt hexagonal symmetry with different edge-termination, armchair (HAC) and zig-zag (HZZ). The  defect level related to the $\sigma$-dangling bond (see Fig. \ref{isosurfs}) shows a value of spin-splitting close to that found for the SC models, causing a magnetic moment of $1.0\mu_B$; moreover, in both cases we find a defect-related $V_\pi$ state, also spin-split, summing a final integer magnetic moment of $2.0\mu_B$. It is interesting to compare as shown in Fig. \ref{isosurfs} the density distribution for the defect-related states $V_\pi$ in the two clusters and in the $(7\times7$) SC, and the $l_\pi$ state of the ($6\times6$) SC. We see clearly the trigonal symmetry in all cases, and also that in the  $(7\times7$) periodic arrangement there is little overlap between the $V_\pi$ states in neighbor cells. In the case of the $l_\pi$ state, we see the $\pi$-connection along the zig-zag lines. Only in the ($6\times6$) SC we have the fully delocalized states $d'_\pi$ contributing to the final spin density.

Up to now, we have conflicting results for the magnetic moment of the same defect, coming from different theoretical modeling (cluster and SCs) based on the same computational code and numerical settings.

At this point, we look at the effect of inclusion of XC in the functional, shown in the results obtained with PBE0 functional.  In the case of the hexagonal clusters the actual value of $\mu_{V}$ does not change $\mu_{V}=2\mu_{B}$, and we see in Fig.\ref{cluster_results} that the main impact is the spin-splitting found for the defect levels, that for the \textit{V}$\pi$ state goes from $\sim{~0.2}$eV to $ 1.3$eV. It is to be noted however that using the PBE0 $\alpha$-fraction we observe, for both clusters, that the defect  level $V\pi$  is now a ``midgap'' state  pinned at $E_{F}=0$, as seen by Ugeda et. al.~\cite{uged+10prl} and in accordance with previous theoretical predictions~\cite{pere+08prb}, which is not the case using PBE.  We pass next to the more impactant effect, seen for  all SCs and shown in Fig.\ref{v_level_band_PBE_PBE0_sp} (b): we find that inclusion of XC eliminates the band-crossing at the Fermi energy, enhancing the spin-splitting for the involved states and restoring the vacancy magnetic moment, $\mu_{V}=2\mu_{B}$.

Even if the magnetic moment is now the same, still for the ($6\times6$) SC it comes from the splitting of the \textit{d'}$\pi$ levels, not from the defect-localized states, while in the case of the other two SCs the integer magnetic moment comes from the complete spin-splitting of the \textit{V}$\pi$ state. Indeed, from the ($7\times7$) to the ($8\times8$) SC both acceptor \textit{V}$\pi_\uparrow$ and donor \textit{V}$\pi_\downarrow$  levels approach the Fermi energy, but showing a different localization character (band curvature close to the $K_{SC}$ point)  as detected in experimental results~\cite{zhan+16prl}.

\begin{figure}
		\centering
			 \includegraphics[width=0.5\textwidth]{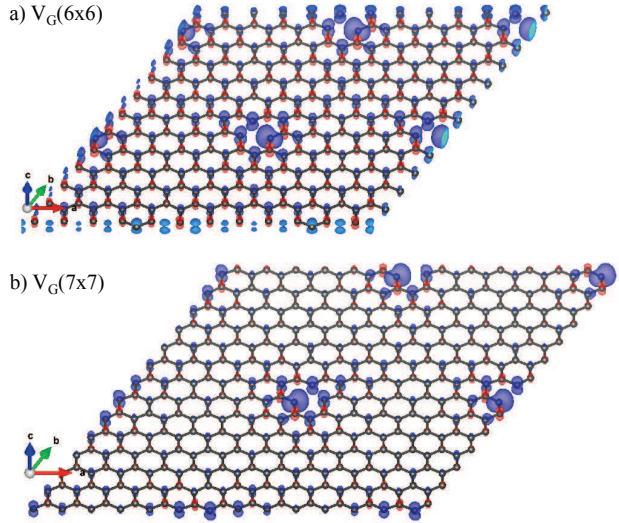}
		\caption{(Color online)Isosurfaces for the spin density ($0.05\AA{}^{-3}$) produced by the array of vacancies in graphene  obtained through PBE0 at the $\Gamma_{SC}$ point for(a) $6\times6$ and (b) $7\times7$ SCs. }
		\label{vSC_PBE0_SD}
\end{figure}

Grouping our results from cluster and periodic boundary conditions, we see that with the inclusion of XC we can predict an integer magnetic moment of $\mu_{V}=2\mu_{B}$ for the isolated vacancy defect. The characteristic \textit{V}$\sigma$ level, seen in different DFT studies, shows a large spin splitting of very similar magnitude in our different simulations. For the defect $\pi$-states, we also see a characteristic acceptor level in the cluster and ($3n\pm1$) cells, pinned to the Fermi energy, responsible for the final integer magnetic moment. The confinement effect in the cluster models place the donor level much below, however from periodic conditions, in the ($3n\pm1$) cells, we see this level approaching the Fermi energy.

We turn now to the specific results obtained for the ($6\times6$) SC: the plot in Fig. \ref{vSC_PBE0_SD}, showing the \textit{spin density} across the cell, highlights the delocalized effect of this  $3n$-array of defects compared to the immediately one-unit larger ($7\times7$) SC. The high spin-density centered on the vacancy site comes from the difference in density between the $l_\pi$ up and down states, while the overall delocalization comes from the mixed $d'\leftrightarrow l$ character. We suggest this symmetry-derived behavior could be explored by designing chosen arrays of point defects.

	\begin{acknowledgments}

This work was supported by FAPESP, INEO, and CAPES, Brazil and CONICYT, Chile. MJC acknowledges support from CNR-S3, Italy. We also acknowledge support on computer time  by the Res. Comput. Support Group (Rice University) and LCCA-USP, and NAP-NN-USP. We thank E. Molinari and L.G Dias da Silva for fruitful discussions.

	\end{acknowledgments}
	
\bibliography{names,Vac_Graph_Biblio}
	
\end{document}